\documentclass[twocolumn]{aastex7}

\begin{document}

\title{Calibrated Period-Luminosity Relations for Mira variables in the Large Magellanic Cloud in the Vera C. Rubin Observatory $zy$-band Photometric Systems}

\author[0000-0001-8771-7554,gname=Chow-Choong, sname=Ngeow]{Chow-Choong Ngeow}
\affil{Graduate Institute of Astronomy, National Central University, 300 Jhongda Road, 32001 Jhongli, Taiwan}
\email[show]{cngeow@astro.ncu.edu.tw}

\author[0000-0001-6147-3360,gname=Anupam, sname=Bhardwaj]{Anupam Bhardwaj}
\affil{Inter-University Center for Astronomy and Astrophysics (IUCAA), Post Bag 4, Ganeshkhind, Pune 411 007, India}
\email{anupam.bhardwajj@gmail.com}


\begin{abstract}

The Vera C. Rubin Observatory's Legacy Survey of Space Time (LSST) is expected to discover numerous Mira variables in nearby galaxies, enabling their application for the cosmic distance scale. Miras are excellent distance indicators at infrared wavelengths, and therefore their period-luminosity (PL) relations in reddest $zy$ photometric systems will be crucial in the LSST era. We present first empirical $zy$-band PL relation for 85 oxygen-rich (O-rich) Miras located in the Large Magellanic Cloud based on time-series observations collected using the Dark Energy Camera (DECam) at the Víctor M. Blanco 4-m Telescope. The 
DECam photometry was directly calibrated using the Rubin Observatory Data Preview 2 (DP2) data. The observed PL relations were found to exhibit a break around a period of $\sim 300$~days, and therefore we only derived and calibrated the PL relations for Miras with a period shorter than $300$~days. We also explored the Rubin LSST DP2 data for known Mira variables in three Local Group dwarf galaxies. The $zy$-band light curves of Miras in IC 1613 are too sparsely sampled, whereas the DP2 observations of NGC 6822 and the Sagittarius Dwarf Irregular Galaxy cover only a short temporal baseline. For an O-rich Mira in NGC 6822, we obtain a distance modulus $\mu = 23.32\pm0.24$~mag in good agreement with the literature. Significantly improved temporal coverage is expected in the first official Rubin LSST data release that will enable a robust validation of the PL relation in these nearby systems.

\end{abstract}

\section{Introduction}

The Mira variables (hereafter Miras) are long-period pulsating stars with a period ranging from $\sim100$ to $\sim1000$~days. As late-type cool asymptotic-giant-branch (AGB) stars, the spectral-energy distribution (SED) for Miras peaks around the $K$-band, making this filter an excellent choice for constructing the period-luminosity (PL) relation for distance scale applications. Indeed, most of the Mira's PL relations derived in the literature are in the $K$-band or similar near-infrared (NIR) wavelengths \citep[for example,][and reference therein]{bhardwaj2019, ita2021, sun2022, sun2023, lewis2023, sanders2023, konchady2024, bhardwaj2025}.

As argued in \citet{whitelock2013} and \citet{huang2024}, Miras can be competitive to classical Cepheids for distance scale applications, such as providing independent measurements of the Hubble-Lemaitre constant $H_0$ \citep[for example, see][for the application of Miras found in M101]{huang2024a}. Recently, \citet{bhardwaj2025} provided NIR calibration of Miras in the Galactic globular clusters and used those as an independent anchor to determine a 3.7\% measurement of $H_0$. One advantage of using Miras is that these intermediate to old age stars are found in both spiral and elliptical galaxies, in contrast to young Cepheids which can only be found in spiral galaxies. For general reviews of Miras and their applications to the cosmic distance scale, the readers may refer to \citet{huang2024} and \citet{skowron2026}.

Beginning 30 June 2026, the NSF-DOE Vera C. Rubin Observatory Legacy Survey of Space and Time \citep[hereafter Rubin LSST,][]{lsst2019} will routinely observe the southern sky in six $ugrizy$ filters for a decade. As one of the scientific targets within the Rubin LSST Transients and Variable Star science collaboration \citep{hambleton2023}, numerous ($\sim10^4$ to $\sim 10^5$) Miras are expected to be discovered in nearby galaxies \citep{macri2017}. Clearly, Rubin LSST lacks a $K$-band filter, and its closest or longest wavelength filter for Rubin LSST is the $y$-band. Therefore, it is desirable to derive and calibrate the PL relation for Miras in the $y$-band, such that these empirical calibrations can be applied to the Rubin LSST-discovered Miras for distance scale applications. We note that there is no empirically calibrated $y$-band PL relation for Miras available in the literature, except for the synthetic PL relations presented in \citet{iwanek2021}.

\begin{figure*}
  \epsscale{1.15}
  \plotone{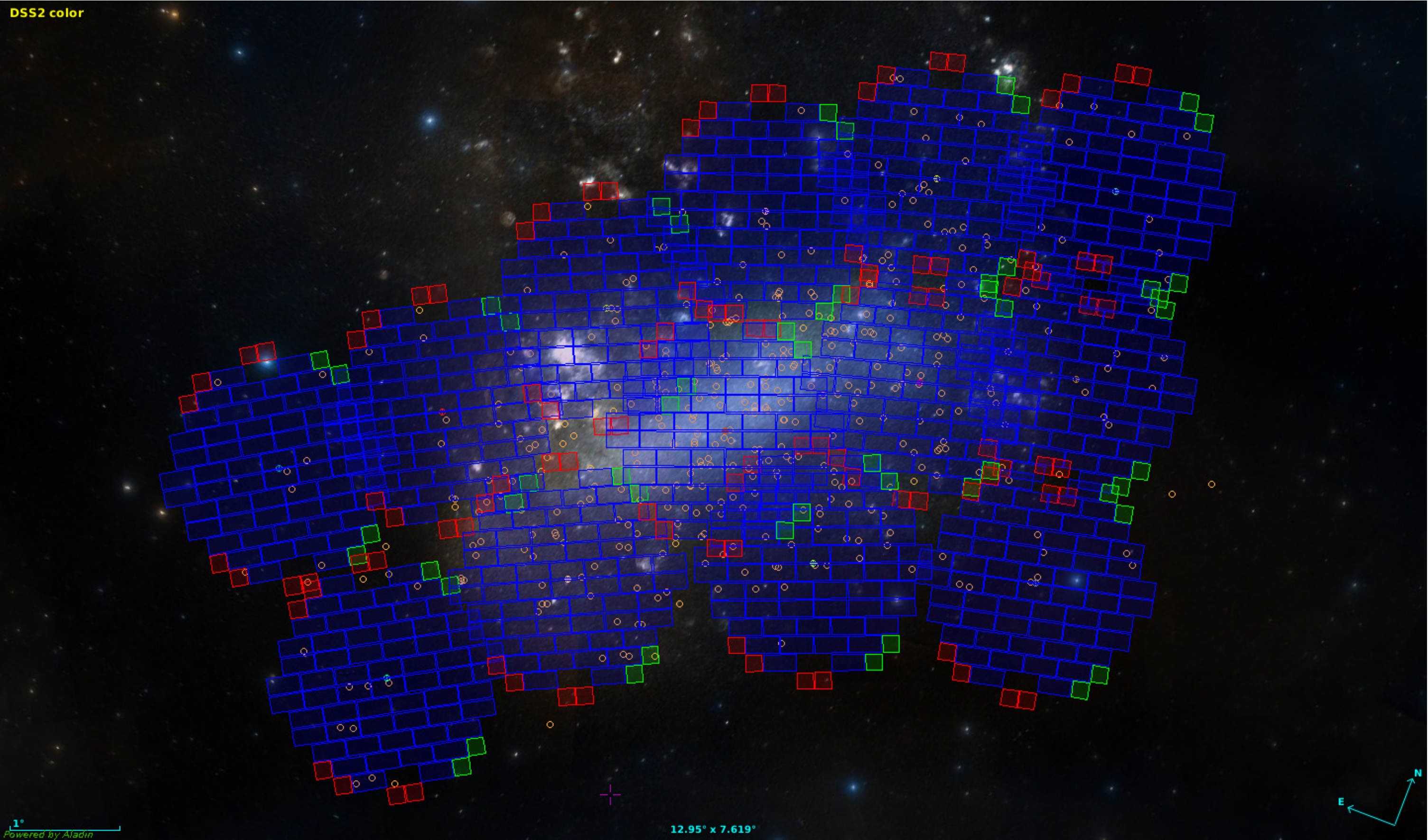}
  \caption{The distribution of 13 DECam pointings toward LMC to cover the selected O-rich LMC Miras \citep[in yellow circles, selected from][]{ou2022} as much as possible. The green and red squares represent the guiding and focusing CCD chips on the DECam, respectively. This plot was prepared using the {\tt Aladin} software \citep{bonnarel2000}.}
  \label{fig_lmc}
 \end{figure*}

As a result, we initiated a project to derive the $y$-band PL relation using the known Miras located in the Large Magellanic Cloud (LMC). This is because LMC has a well-measured distance to the precision of the 1\%~level \citep{pietrzynski2019,lengen2026}, and LMC is located within the visibility area of Rubin LSST. Furthermore, LMC hosts a large population of Miras with well-determined periods from the Optical Gravitational Lensing Experiment \citep[OGLE,][]{ogle2009}, and hence the periods and other pulsation properties are already known. In particular, we adopted a ``clean'' sample of LMC Miras from \citet{ou2022}, composed of oxygen-rich (O-rich) variables\footnote{We do not consider the carbon-rich (C-rich) Miras in this work.} with well-behaved (i.e. ``regular'') light curves for the derivation of the $y$-band PL relation. 

Ideally, time-series data obtained from Rubin LSST could provide necessary light curve data sets for the LMC Miras. However, Rubin LSST saturates around $y\sim14$~mag, which is roughly the same as the mean $y$-band magnitude for LMC Miras \citep{iwanek2021}. Clearly, saturated Rubin LSST photometry would affect the light curve quality and hence the derived mean magnitudes. To remedy this, we observed the LMC Miras using the wide-field DECam \citep[Dark Energy Camera,][]{decam2015}, mounted on the Víctor M. Blanco 4-meter Telescope. This was done by taking a series of shallow images so that the LMC Miras were not saturated. Since Miras are intrinsically red, the DECam observations have also covered the $z$-band to include the $(z-y)$ color-term during the photometric calibrations. Because of this, we have also derived the $z$-band PL relation in addition to the $y$-band PL relation.

\section{DECam Observations} \label{sec_obs}

\begin{figure*}
  \epsscale{1.15}
  \plotone{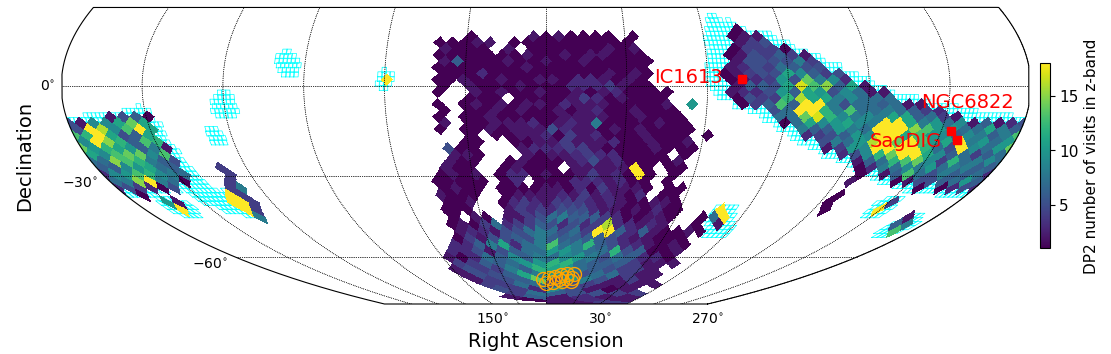}
  \plotone{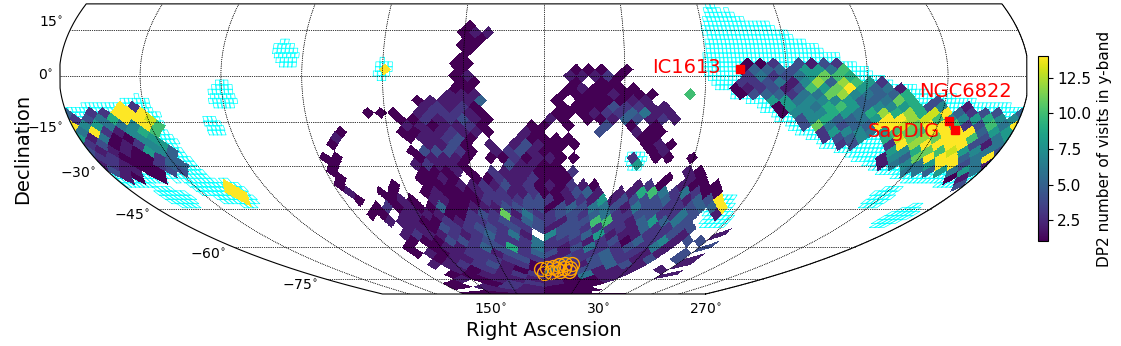}
  \caption{The top and bottom panel display the sky coverage of DP2 in the $z$-band and $y$-band, respectively. The colorbars represent the number of DP2 visits on each LSSTCam observations. The 13 DECam pointings on LMC are marked as orange circles, along with the locations of three dwarf galaxies investigated in Section \ref{sec_dp2} shown as red symbols. The cyan rectangles represent the DP2 coadded tracts.}
  \label{fig_dp2}
 \end{figure*}

Given the large field-of-view (FOV) for DECam, about 3 square degrees, most of the target $\sim 400$ regular O-rich LMC Miras can be covered within 13 DECam pointings (see Figure \ref{fig_lmc}). A sequence of $zy$-band images with exposure time of $5$ and $10$~seconds, respectively, were taken for each of the DECam pointings to avoid saturation in the LMC Miras. With such exposure time in the $zy$-band, together with overheads, the observation time to cover 13 DECam pointings would take about $20$~minutes in a given night. This 20~minute observation represents one epoch on the light-curve, and multiple epoch observations were carried out in the NOIRLab 2024B and 2025B Semesters. Because such a short observation time is needed per night but is repeated in multiple nights, our ``short-and-repeat'' remote observation experiments were carried out within the DECam Alliance for Transients (DECAT) consortium \citep[for a further description, see][]{graham2023}.

Due to multiple constraints, including the availability of the $y$-band filter on DECam, the nights awarded for the DECAT consortium, priority of the targets within the DECAT pool, etc., our program was carried out in 11 and 3 nights in the 2024B and 2025B Semesters, respectively, by the DECAT remote observers. The 11 nights in the 2024B Semester include 4 nights in September 2024 and 7 consecutive nights in January 2025. In the 2025B Semester, our program was carried out one and two nights in August and September 2025, respectively. Note that on 14-16 January 2025, the observations were repeated in these three nights. Altogether, a total of 17 epochs of observations have been taken for our program. For a small number of Miras, the number of epochs could be as much as 34, because they are located at the overlapped regions of two DECam pointings.

\section{Photometric Calibration} \label{sec_calib}

\subsection{Preparing the DECam Images} \label{det_img}

All DECam observed images were processed via the DECam community pipeline \citep{valdes2014}, and processed images with {\tt InstCal} types were downloaded for photometric measurements. To save disk space, we have used {\tt MissFITS} \citep{marmo2008} to split the DECam multiextension FITS images into individual CCD chips, and only the CCD chips that contain our target LMC Miras have been kept. Using {\tt Swarp} \citep{bertin2002}, we further median combined all CCD chips in the same DECam pointings with the same CCD numbers to create a deep detection image for a given DECam pointing and CCD chip combination. These processed images are used in the following subsection.

\subsection{Preparing the Reference Catalog}

Prior to the official data releases of the Rubin LSST, the Vera C. Rubin Observatory has released two datasets based on the observations taken during the on-sky commissioning campaign and the early optimization phase, known as Data Preview 1 \citep{dp1} and Data Preview 2 \citep[hereafter DP2,][]{RTN-115,dp2data}.\footnote{The DP1 did not cover LMC and therefore not considered in this work. The DP2 was split into the early DP2 and full DP2 releases, where the early DP2 only includes the coadd images and all of the catalog products. In contrast, the full DP2 will also include the visit-level images. Since in this work we only used the catalog products, we simply refer the early DP2 as DP2.} As shown in Figure \ref{fig_dp2}, LMC is included in the $zy$-band sky coverage of DP2, providing a natural reference catalog to calibrate the DECam photometry.

However, as can be seen from Figure \ref{fig_dp2}, there are no coadded images at the location of LMC, implying that there are no data in the DP2's {\tt Object} table. Therefore, we queried the {\tt Source} table, through the Rubin Science Platform (RSP), to obtain the $zy$-band photometry for the point-like sources (selected using the {\tt extendedness = 0} condition) located within the DECam pointings. Furthermore, we also excluded sources with any of the following flags set to ``TRUE'' in the {\tt Source} table: invalidPsfFlag, pixelFlags\_bad, pixelFlags\_cr, pixelFlags\_edge, pixelFlags\_interpolated, pixelFlags\_nodata, pixelFlags\_offimage, pixelFlags\_saturated, pixelFlags\_suspect. A caveat of using the {\tt Source} table is that a given source would have multiple and unique {\tt sourceID} in this table if it has been observed multiple times (or multiple visits). Therefore, we further split the downloaded {\tt Source} catalogs according to the unique {\tt visit} IDs. 

Using the deep median combined images of the same DECam pointings and CCD chips, prepared in Section \ref{det_img}, we cross-match the split {\tt Source} catalogs with the sources detected in these DECam detection images to build the DP2-LMC reference catalog. For matched sources that have multiple visits in a given filter, we took the weighted average of the available DP2 photometry, else we adopted the single-visit DP2 measurements. However, DP2 did not uniformly cover the entire LMC but with patchy observations, resulting $\sim 5\%$ of the images to have no DP2 sources and $\sim 36\%$ of the images to have fewer than 20 reference sources of DP2. These images were excluded in the subsequent photometric calibration.\footnote{We suspected that the crowding issue is a potential reason for not having sufficient DP2 source detections in these regions. The excluded DECam regions will be revisited when Rubin LSST releases its first official data release in the coming years.}  

\subsection{Running the Photometric Calibration}

For the remaining images, we used the {\tt Source-Extractor} \citep{bertin1996} and {\tt PSFEx} \citep{bertin2011} software suite to measure the instrumental point-spread function (PSF) magnitudes, $m^{PSF}$, in the DP2 reference sources. To ensure that good quality photometry was used in the calibration, we further filtered out those $m^{PSF}$ with errors larger than 0.1~mag and possess {\tt Source-Extractor's flag} values larger than zero. We also excluded sources that exhibit variability in magnitudes observed in two different DP2 visits. We kept those images that have 20 or more DP2-LMC reference sources after such filtering. The photometric calibration involved iteratively solving the following equations \citep[following the same approach as done in][]{ngeow2025}, separately on the $zy$-band images, using the final selected DP2-LMC reference sources:

\begin{eqnarray}
  z^{LSST} - z^{PSF} & = & ZP_z + C_z (z - y)^{LSST}, \label{eqn_z} \\
  y^{LSST} - y^{PSF} & = & ZP_y + C_y (z - y)^{LSST},  \label{eqn_y}
\end{eqnarray}

\begin{figure*}
  \epsscale{1.1}
  \plottwo{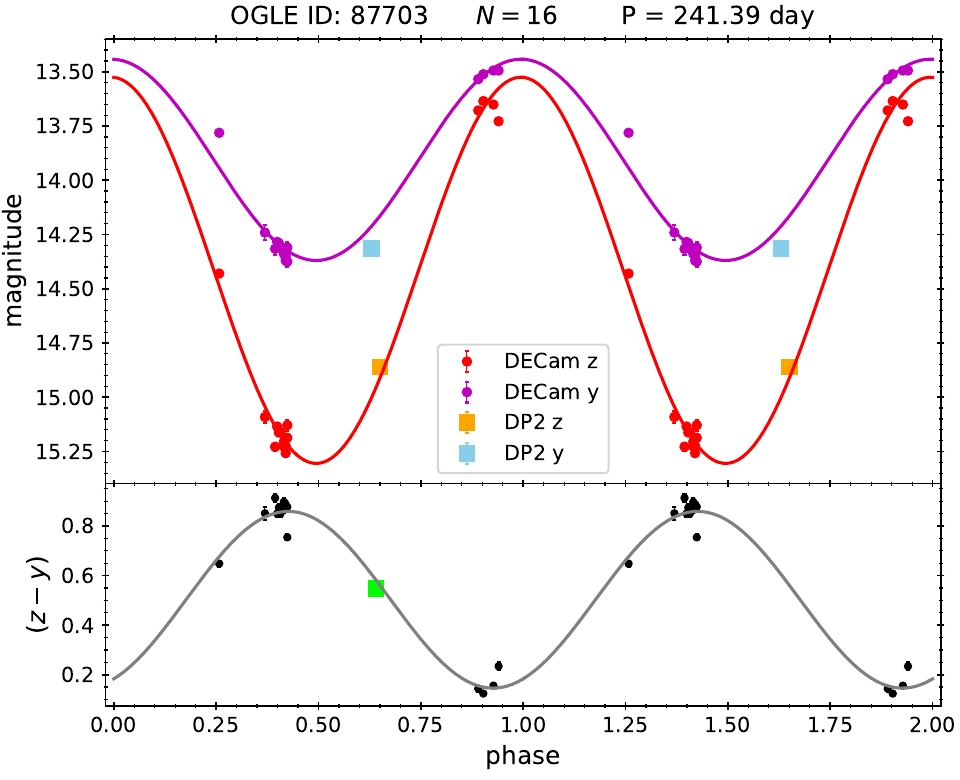}{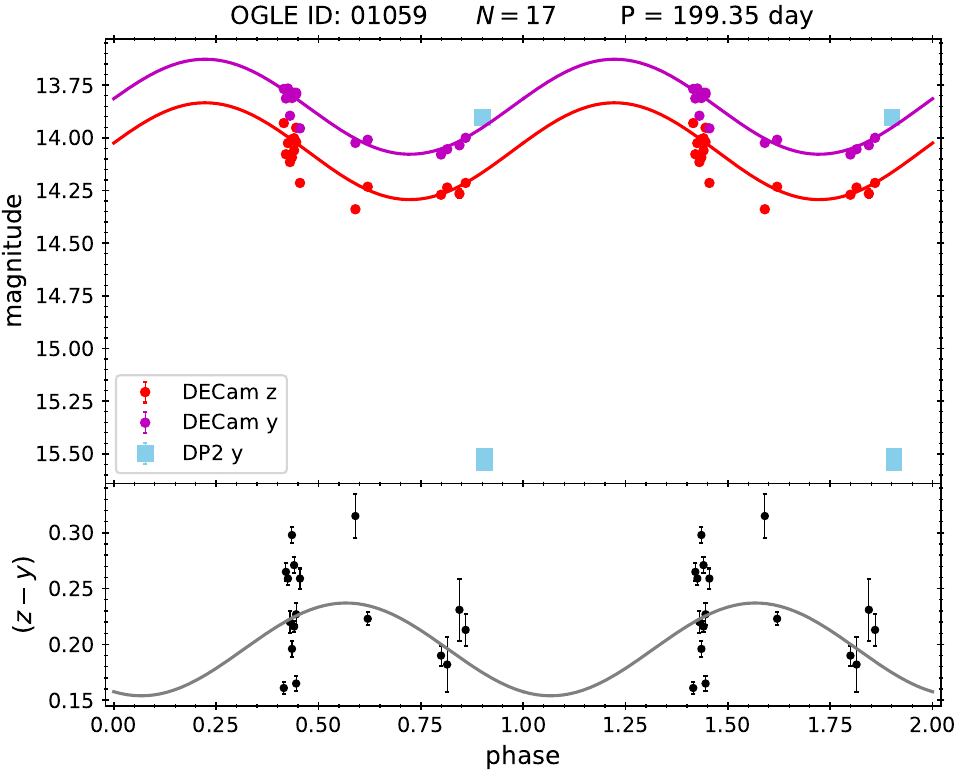}
  \caption{Examples of the $zy$-band light curves and $(z-y)$ color curves for two LMC Miras observed with DECam. The OGLE ID, numbers of data-points ($N$), and the OGLE periods are given on top of each panel. The solid curves are the fitted light curves or color curves using $m=\langle m\rangle+A\sin (2\pi \Phi + \phi)$, where $\langle m\rangle$ is the mean value and $\Phi$ represents the pulsations phases from 0 to 1. The phase shift $\phi$ is forced to be the same for both of the $zy$-band light curves (but not for the color curve). For these two LMC Miras, we have also included the DP2 photometry, queried from the {\tt Source} table and linked the matched sources to the {\tt Visit} table for obtaining the modified Julian date (MJD), as squared symbols.}
  \label{fig_lc}
\end{figure*}

\noindent where $ZP_m$ and $C_m$ represent the zero-points and color-terms in filter $m=\{z,\ y\}$, respectively. The colors in the DP2-LMC reference sources ranged from $\sim -1.5$~mag to $\sim 2.3$~mag.

The mean values of the color-terms on these images are $\langle C_z \rangle=0.534$ and $\langle C_y \rangle=0.100$, with a dispersion of $\sigma_z=0.204$ and $\sigma_y=0.217$, in the $zy$ bands, respectively. We have further removed a small fraction of these images that have a color-term deviating more than $3\sigma$ from the mean values, as the calibrated photometry from these images becomes outlier epochs in the light curves.

\section{Light Curves and Color Curves} \label{sec_lc}

For the photometrically calibrated images, we extracted the instrumental PSF magnitudes for those LMC Miras located within these images in the same way as the reference stars, and calibrated their photometry via equations \ref{eqn_z} and \ref{eqn_y}. Since a pair of $zy$-band images was taken near simultaneously, which allowed the $(z-y)^{LSST}$ color for a given Mira to be solved:

\begin{eqnarray}
  (z - y)^{LSST} & = & \frac{(ZP_z + z^{PSF}) - (ZP_y + y^{PSF})}{1 - C_z - C_y}. \label{eqn_c}
\end{eqnarray}

\noindent The solved color, $(z - y)^{LSST}$ ranges from $\sim 0.0$~mag to $\sim 1.3$~mag, was then applied to equations \ref{eqn_z} and \ref{eqn_y} to calibrate the $zy$-band magnitudes for a given LMC Mira. Using such an approach, we can construct not only the $zy$-band light curves but also the color curves. During the construction of the light and color curves, we have also excluded a pair of $zy$-band magnitudes if either one of them has a photometric error larger than 0.1~mag (they also tend to be outliers in the light and/or color curves). Therefore, a given Mira would have the same number of data-points ($N$) in both of the $zy$ bands. Two examples of the $zy$-band light curves and color curves are presented in Figure \ref{fig_lc}. We further removed those Miras that have five or less data-points left on the light and color curves, as they are too sparse to have a meaningful analysis. 

After applying various selection criteria, as mentioned previously, there were about 100 LMC Miras left in our sample. The published pulsation periods ($P$) were used to fold their $zy$-band light curves and the color curves. Given the sparse nature of our light/color curves, we fit these phased curves using a sinusoidal function to determine the mean magnitudes. We visually inspected the fitted light and color curves, and excluded those Miras that exhibit unrealistic large amplitudes. This happened in cases where data-points clustered in a narrow range of the pulsation phases. As a result, our final sample consists of 85 LMC Miras.

We have also queried the DP2 $zy$-band light curves, via the RSP on the {\tt Source} table, for our entire LMC Mira sample. However, when using a search radius of $0.5\arcsec$, only about half of them have at least one measurement in the $z$- or $y$-band.\footnote{We have also experimented with an enlarged search radius of $1\arcsec$, but this would also include many spurious sources.} Furthermore, the DP2 photometry seems to agree with our fitted light curves only in very few cases. An example is shown in the left panel of Figure \ref{fig_lc}, where the $zy$-band photometry queried from the DP2's {\tt Source} table was located very close to the fitted curves. In many cases, the DP2 photometry is either too faint or too bright. An example is displayed in the right panel of Figure \ref{fig_lc} -- even though there is one data-point that agrees with the fitted curve, there are two outlier data-points at $y\sim 15.5$~mag. Since the LMC Miras are quite bright for the Rubin Observatory, it is possible that DP2 photometry is not optimized for bright sources yet. Therefore, we did not include the DP2 photometry when fitting the sinusoidal curves.

\begin{deluxetable*}{cccrccc}
  \label{tab1}
  \tabletypesize{\scriptsize}
  \tablecaption{Mean $zy$-band Magnitudes and Colors for the LMC Miras.}
  \tablewidth{10pt}
  \tablehead{
    \colhead{OGLE ID} &
    \colhead{$P$ (days)} &
    \colhead{$E(V-I)$} &
    \colhead{$N$} &
    \colhead{$\langle z\rangle$} &
    \colhead{$\langle y\rangle$} &
    \colhead{$\langle (z-y)\rangle$} 
  }
  \startdata
OGLE-LMC-LPV-00743&216.37&0.179&16&$14.009\pm0.002$&$13.776\pm0.003$&$0.292\pm0.003$ \\
OGLE-LMC-LPV-00902&296.25&0.177&16&$13.793\pm0.010$&$13.313\pm0.012$&$0.473\pm0.007$ \\
OGLE-LMC-LPV-00990&321.41&0.124&17&$13.081\pm0.004$&$12.859\pm0.003$&$0.254\pm0.005$ \\
OGLE-LMC-LPV-01059&199.35&0.117&17&$14.063\pm0.003$&$13.852\pm0.003$&$0.196\pm0.006$ \\
OGLE-LMC-LPV-01199&132.81&0.062&17&$15.010\pm0.003$&$14.732\pm0.003$&$0.296\pm0.005$ \\
  $\cdots$&$\cdots$&$\cdots$&$\cdots$&$\cdots$&$\cdots$&$\cdots$ \\
\enddata
\tablecomments{This table is available in its entirety in machine-readable form. A portion is shown here for guidance regarding its form and content.}
\end{deluxetable*}

\section{The PL Relations} \label{sec_pl}

In Table \ref{tab1}, we summarize the mean magnitudes and mean colors for our final sample of LMC Miras. These mean magnitudes and mean colors were derived based on the fittings of a sinusoidal curve (see the caption in Figure \ref{fig_lc}). In the same Table, we also included the values of $P$ and $E(V-I)$ adopted from \citet[][and reference therein]{ou2022}. For extinction and reddening corrections, we adopted $R_z=1.323$ and $R_y=1.088$ \citep{schlafly2011}, together with a conversion of $E(B-V)=E(V-I)/1.237$ \citep{skowron2021}. 

\begin{figure}
  \epsscale{1.1}
  \plotone{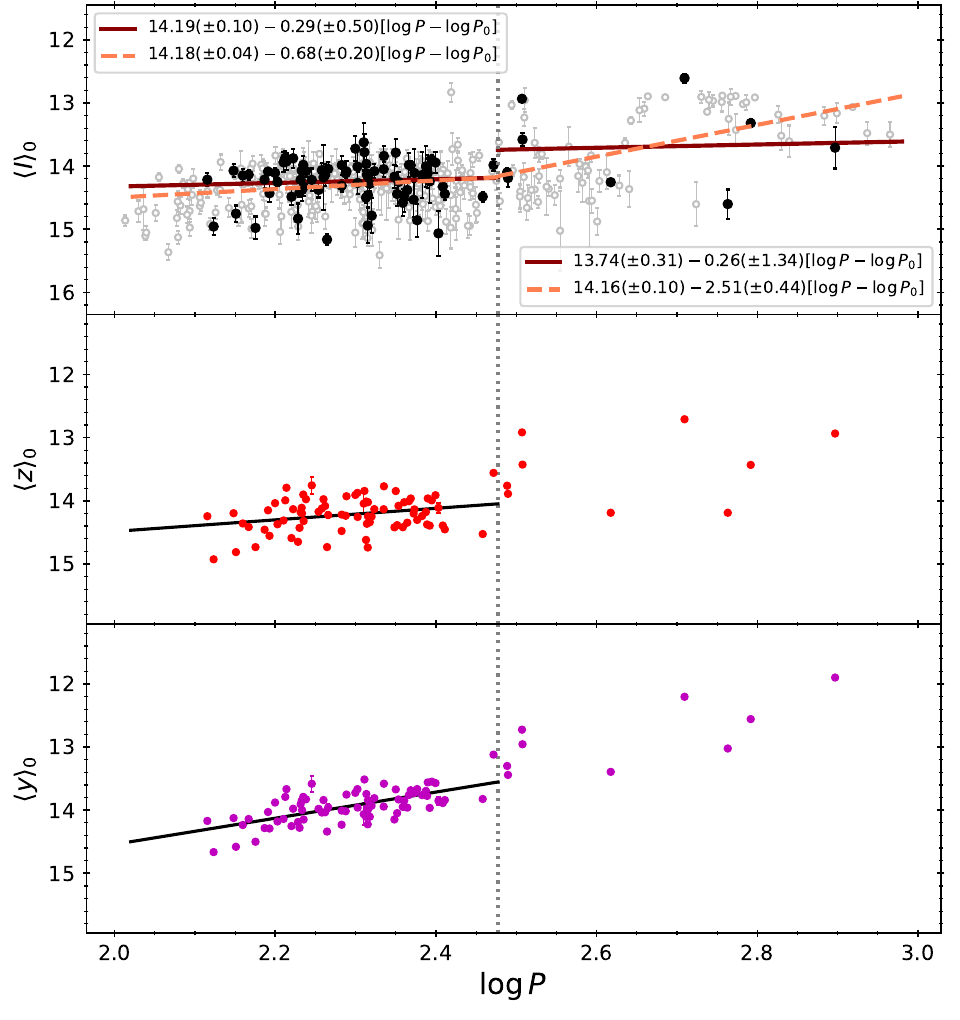}
  \caption{{\it Upper Panel:} Comparison of the $I$-band PL relation for the full sample of LMC O-rich Miras as presented in \citet[][open gray circles]{ou2022} and our final selected sample (filled black circles). The dashed and solid lines are the fitted PL relations to the full and our selected samples, respectively. {\it Middle and Lower Panels:} The $zy$-band PL relations for our final selected sample. The solid lines are the fitted PL relations to the short period Miras, as given in equation \ref{eqn_plz} and \ref{eqn_ply}. In all panels, the mean magnitudes have been corrected for extinction. The vertical dotted line indicates the adopted break period of $P_0=300$~days.}
  \label{fig_pl}
\end{figure}

In the upper panel of Figure \ref{fig_pl}, we compare the (extinction-corrected) $I$-band PL relations between the full sample adopted from \citet{ou2022} and our final selected sample. As can be seen from the entire sample, the $I$-band PL relation exhibits a break around $\log P\sim 2.4$ to $\sim 2.5$. By adopting a break period at $P_0=300$~days \citep{bhardwaj2019}, we fit the $I$-band PL relation separately for the short ($P<P_0$) and long ($P>P_0$) period Miras in the form of $I=a_I + b_I(\log P - \log P_0)$.

When we compared the $I$-band PL relations fitted for the full sample and our final selected sample, we found that the PL relations derived for the short period Miras are in good agreement. This suggests that our final selected sample can be used to derive the short period $zy$-band PL relations, as given below:

\begin{figure*}
  \epsscale{1.1}
  \plottwo{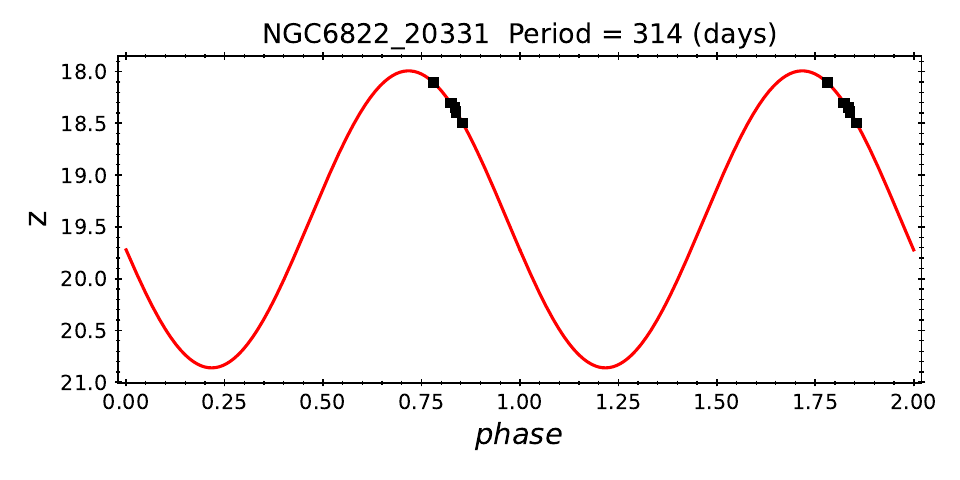}{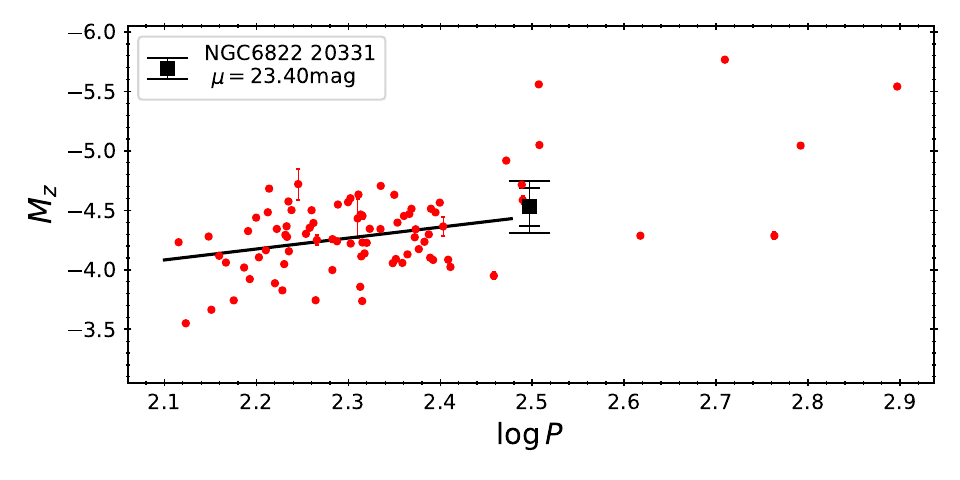}
  \caption{{\it Left Panel:} The DP2 light curve data (black symbols) for an O-rich Mira, with ID 20331 \citep{whitelock2013b}, in NGC~6822. The red curve is the fitted sinusoidal curve as done in Figure \ref{fig_lc}. {\it Right Panel:} Comparison of the $z$-band absolute magnitude for Mira 20331 (black symbol), after correcting for extinction using the \citet{sfd1998} extinction and adopting $\mu_{6822}=23.40$~mag (see text for details), with the LMC O-rich Miras (red circles). A distance modulus of $\mu_{LMC}=18.48$~mag \citep{pietrzynski2019} was used to convert the observed mean magnitudes of LMC Miras to the absolute magnitudes, and calibrate the PL relation derived in equation \ref{eqn_plz} (solid line). For the NGC~6822 Mira, the shorter error bar represents the maximum and minimum values of $\mu_{6822}$ listed in \citet{parada2023}, and the longer error bar is for the $0.22$~mag error in the mean magnitude.} 
  \label{fig_6822}
\end{figure*}

\begin{eqnarray}
  \langle z\rangle_0 & = & 14.05(\pm0.09) - 0.92(\pm0.45)~ [\log P - \log P_0], \label{eqn_plz} \\
  \langle y\rangle_0 & = & 13.55(\pm0.06) - 2.07(\pm0.34)~ [\log P- \log P_0], \label{eqn_ply} 
\end{eqnarray}

\noindent with a dispersion of $\sigma_z=0.26$~mag and $\sigma_y=0.20$~mag, respectively. However, discrepancies in the PL relations were found for the long period Miras (as shown in the upper panel of Figure \ref{fig_pl}). Furthermore, there are only a few of the long period LMC Miras left in our sample, and they exhibit a larger dispersion than their short period counterparts. Therefore, we did not derive the long period $zy$-band PL relations.

\section{Dwarf Galaxies Miras Observed in DP2} \label{sec_dp2}

Since the short-period O-rich Miras in the Galactic field \citep[e.g.][]{sanders2023} and globular clusters \citep[e.g.][]{bhardwaj2025} are too bright for the Rubin Observatory, the best way to test our derived PL relations is using Miras in the local dwarf galaxies. There are three dwarf galaxies located in the DP2 footprint (see Figure \ref{fig_dp2}) with Miras detected in the past. These three dwarf galaxies are IC~1613 \citep{menzies2015}, NGC~6822 \citep{whitelock2013b}, and the Sagittarius dwarf irregular galaxy \citep[hereafter SagDIG,][]{whitelock2018}, providing a great opportunity to test our derived PL relations directly with the DP2 data. 

Using the LSDB framework \citep[Large Scale Database,][]{caplar2025} within the RSP, we queried the DP2 {\tt DiaObject} catalog and the nested {\tt diaObjectForcedSource} tables to extract the $zy$-band light curves for the O-rich Miras. The post-processing of the LSDB framework has also converted the PSF fluxes given in the DP2 tables to AB magnitudes \citep[for details, see][]{malanchev2025}, providing a convenient way to construct the light curves. When constructing the light curves, we have also excluded data-points with any of the following flags set to {\tt True}:  psfFlux\_flag, pixelFlags\_bad, pixelFlag\_saturated, pixelFlags\_cr, pixelFlags\_nodata, pixelFlags\_suspect, pixelFlags\_edge, and pixelFlags\_interpolated.

\subsection{NGC~6822}

NGC~6822 has the highest number of known Miras among the three dwarf galaxies, with 61 Miras detected in \citet{whitelock2013b}. The 11 O-rich Miras in NGC~6822 have 7 to 11 DP2 data-points available on their $zy$-band light curves. However, DP2 $y$-band observations on the location of NGC~6822 only spanned $\sim 3.9$~days. This implied that the $y$-band light curves data, after folded with Mira's period, would cluster in a narrow range of the pulsation phases and hence it would be impractical to fit a sinusoidal curve. The data for the $z$-band light curves spanned a longer observing time of about $23.1$~days. The ranges of the pulsation phases are slightly wider than in the $y$-band; however, the fitted sinusoidal curves still returned unrealistic mean magnitudes and/or amplitudes for most of the Miras in NGC~6822. The only exception is Mira with ID 20331, which has a fitted mean magnitude of $\langle z \rangle=19.15\pm0.22$~mag and an amplitude of $2.87$~mag. Its DP2 $z$-band light curve and the fitted sinusoidal curve are shown in the left panel of Figure \ref{fig_6822}.

To compare this NGC~6822 Mira with the LMC Miras, we adopted a distance modulus of $\mu_{6822}=23.40$~mag to NGC~6822 and converted the extinction-corrected mean $z$-band magnitude of this Mira to absolute magnitude. The value of $23.40$~mag is the mid-point of the measurements collected in Table 10 of \citet[][spanning from 23.24 to 23.56~mag]{parada2023}. The right panel of Figure \ref{fig_6822} compares the absolute magnitude of these Miras to those of the LMC, showing that this NGC~6822 Mira is located within the parameter space defined by the LMC Miras. 

Alternatively, we can evaluate the performance of our derived PL relation by deriving the distance modulus to NGC~6822. Since this Mira has a period of $314$~days, slightly longer than $P_0$, if we assumed that the extrapolated $z$-band PL relation is valid for this Mira, then we obtained a distance modulus of $\mu_z = 23.32\pm0.24$~mag, which is consistent with the values compiled in \citet{parada2023}. 

\begin{figure*}
    \epsscale{1.15}
    \plottwo{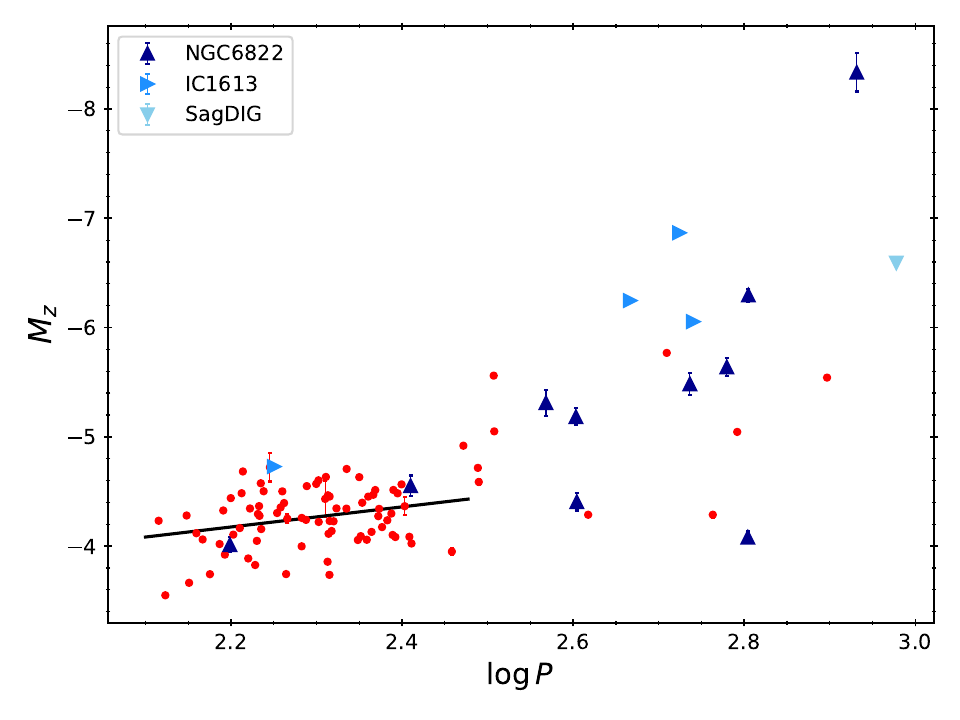}{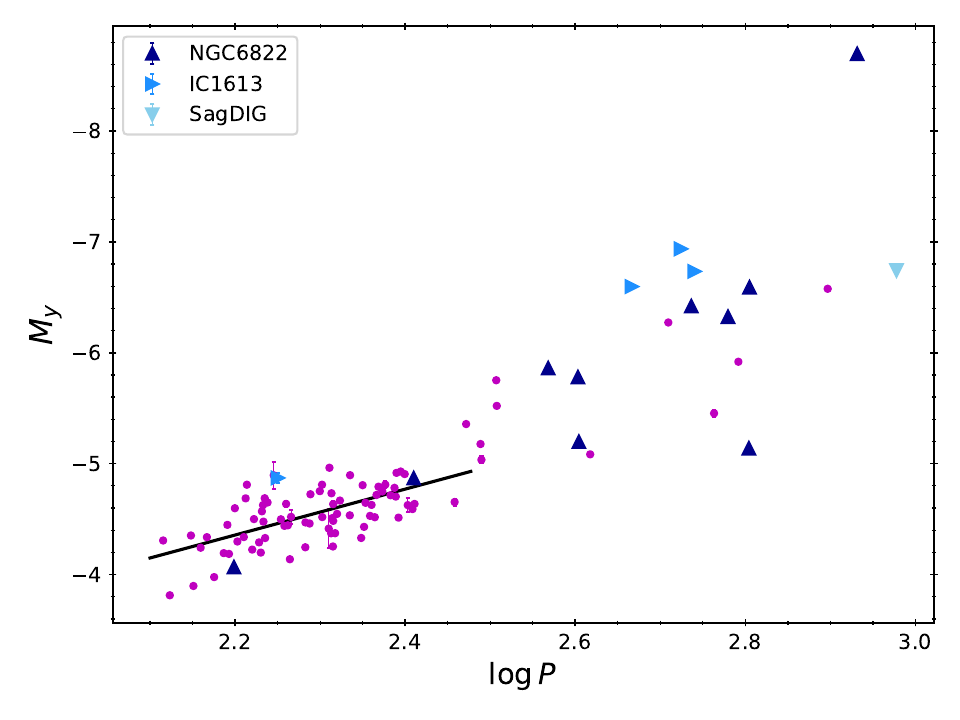}
    \caption{The PL relations in the $zy$ bands for Miras similar to the right panel of Figure \ref{fig_6822}. O-rich Miras in NGC~6822, IC~1613, and SagDIG are shown along with the LMC Miras. The extinction-corrected "magnitudes" for the Miras in the dwarf galaxies are based on the equivalent single-epoch random-phase measurements (see text for details). Same as NGC~6822, we adopted the distance modulus of IC~1613 as the mid-point of the values listed in Table 11 of \citet{parada2023}, as $\mu_{1613}=24.34$~mag. The distance modulus of SagDIG, $\mu_{SagDIG}=25.27$~mag, was adopted from \citet{parto2023}.}
    \label{fig_cmp}
\end{figure*}

For the remaining 10 O-rich Miras that are not fitted with a sinusoidal curve, we took a weighted mean of their light curve data and treated it as a single-epoch random-phase magnitude. In Figure \ref{fig_cmp}, we compare such single-epoch random-phase magnitudes, after converting those to absolute magnitudes, to the LMC Miras in both $zy$-band PL relations. As expected, they show a larger scatter, especially for the long period Miras, due to incompleteness of phase sampling. Adding longer baseline light curve data from the Rubin LSST would naturally improve the measurements of their mean magnitudes, and hence reduce the scatter on the PL relations.

\subsection{SagDIG and IC~1613}

The only O-rich Mira in SagDIG reported by \citet{whitelock2018} has a very long pulsation period of 950~days and is undergoing a hot-bottom burning phase. Given its projected location close to NGC~6822 (see Figure \ref{fig_dp2}), this Miras has a very similar number of DP2 $zy$-band data-points as the Miras in NGC~6822. The observing time-spans, on the other hand, are $\sim 12.3$ and $\sim 37.2$~days on the $z$- and $y$-band light curves, respectively. Given that this Mira has a very long period, the light curve data-points cluster in a narrow range of pulsation phases and are not suitable to fit a sinusoidal curve. As in the case of NGC~6822 Miras, we took a weighted mean for the light curve data of this Mira and treated it as a single-epoch random-phase measurement. Its location in the PL relation is shown in Figure \ref{fig_cmp}.

For IC~1613, 5 of the 11 Miras found in \citet{menzies2015} are O-rich Miras, and the $zy$-band light curve data were available for 4 of them. However, there are only 1 to 2 data-points available on the light curves. Treating them the same way as the Miras in NGC~6822 and SagDIG, we averaged the light curve data if it has two measurements, or adopted the single data-point as the single-epoch random-phase magnitudes. The comparison with the LMC Miras in Figure \ref{fig_cmp} shows large scatter at the long-period end. Therefore, Mira light curves with limited sampling in nearby dwarf galaxies are not very useful for the distance measurement with current data preview releases of Rubin LSST. 

\section{Discussion and Conclusion} \label{sec_last}

In the era of Rubin LSST, the $zy$-band PL relations for Miras would be undoubtedly valuable especially toward the latter half of its decade-long survey when a large number of Mira variables are discovered in numerous galaxies. In this work, we present the first empirical $zy$-band PL relations of a sample of O-rich Miras derived from shallow observation of the LMC with DECam. This is particularly important because the photometry of Rubin LSST is likely to have a saturation problem at the bright end. Indeed, when we examined the DP2 $zy$-band photometry for our sample of Miras, most of the photometric data were found to be spurious, i.e. too faint or too bright.   

Our DECam photometry is directly calibrated using the DP2 data; therefore, these $zy$-band PL relations represent the first empirical study in the Rubin LSST photometric system. A drawback of using DP2 photometry is that the current observations toward the LMC are patchy and incomplete, resulting in a large fraction of our targeted LMC Miras to remain photometrically un-calibrated. Nevertheless, we tested the derived PL relations for short period O-rich Miras using the $I$-band data between the full sample and our final calibrated sample, and both PL relations show a good agreement. Therefore, we derived the $zy$-band PL relations for O-rich Miras with periods shorter than 300~days, after calibrating with the LMC distance \citep{pietrzynski2019}, as:

\begin{eqnarray}
  M_z & = & -4.43 (\pm0.09) - 0.92(\pm0.45)~ [\log P - \log P_0], \nonumber \\
  M_y & = & -4.93(\pm0.07) - 2.07(\pm0.34)~ [\log P- \log P_0]. \nonumber 
\end{eqnarray}

In our work, we assume that the $zy$-band PL relations for the O-rich Miras are insensitive to metallicity, since our DECam observations were not designed to investigate the metallicity effect. However, as pointed out by the referee, a recent study by \citet{huang2026} showed that the spectral properties of O-rich Miras could be affected by metallicity and may have an impact on the applications of the $zy$-band PL relations for stellar systems with different metallicities. \citet{bhardwaj2025} found no significant metallicity effect on $JHK_s$ PL relations for O-rich Miras in star clusters that cover a wide range of mean-metallicities. Therefore, the metallicity dependence of the reddest LSST $y$ band PL relation may not be significantly different from the $J$ band PL relation. A detailed investigation of the metallicity dependence is beyond the scope of this work.

In addition to PL relations, we have also attempted to derive the period-Wesenheit (PW) relation, in which the Wesenheit index is defined as $W = y - 4.630(z-y)$. However, the fitted PW relation carried a large dispersion of $0.42$~mag, which is twice that of the $y$-band PL relation, and it will not be useful for distance measurement. This is likely because the $zy$-band Wesenheit index is very sensitive to color, with a coefficient of $4.63$ in the color term.

There are three dwarf galaxies located within the footprint of DP2, and they are known to host at least one O-rich Miras. This provides a natural testbed for our derived $zy$-band PL relations. However, the time-series $zy$-band observation for IC~1613 is too sparse to fit a light curve. In case of NGC~6822 and SagDIG, which have a higher number of data-points on the $zy$-band light curves, the observations time-spans are too short when compared to the pulsation periods of Miras, and cause the fitted mean magnitudes and/or amplitudes to be unreliable. The only exception is Mira 20331 in NGC~6822, which has a reasonable mean magnitude and amplitude after fitting its $z$-band light curve data with a sinusoidal curve. Nevertheless, by adopting the distance modulus of NGC~6822 from the literature, the absolute $z$-band magnitude for this Mira agrees with LMC Miras with similar periods. We emphasize that this is a preliminary test, because the DP2 data were taken during the commissioning phase. We expect more complete light curve data to be available by combining the DP2 data and the first Rubin LSST data release (DR1) in the coming years.

\begin{acknowledgments}
 The authors thank several useful comments and suggestions of the anonymous referee, which helped improve the manuscript. CCN acknowledges the funding from the National Science and Technology Council (NSTC, Taiwan) under the grants 114-2112-M-008-011 \& 115-2112-M-008-003. AB acknowledges the funding from the Anusandhan National Research Foundation (ANRF) under the Prime Minister Early Career Research Grant scheme (ANRF/ECRG/2024/000675/PMS). This research was supported by the International Space Science Institute (ISSI) in Bern/Beijing through the ISSI/ISSI-BJ International Team project ID $\#$24-603 – ``EXPANDING Universe'' (EXploiting Precision AstroNomical Distance INdicators in the Gaia Universe). We sincerely thank A. Rest, and the observers from the DECAT team for carrying out the remote DECam observations required for our project. This research used Astropy,\footnote{\url{http://www.astropy.org}} a community-developed core Python package for Astronomy \citep{astropy2013,astropy2018,astropy2022}.

  This project used data obtained with the Dark Energy Camera (DECam), which was constructed by the Dark Energy Survey (DES) collaboration. Funding for the DES Projects has been provided by the US Department of Energy, the U.S. National Science Foundation, the Ministry of Science and Education of Spain, the Science and Technology Facilities Council of the United Kingdom, the Higher Education Funding Council for England, the National Center for Supercomputing Applications at the University of Illinois at Urbana-Champaign, the Kavli Institute for Cosmological Physics at the University of Chicago, Center for Cosmology and Astro-Particle Physics at the Ohio State University, the Mitchell Institute for Fundamental Physics and Astronomy at Texas A\&M University, Financiadora de Estudos e Projetos, Fundação Carlos Chagas Filho de Amparo à Pesquisa do Estado do Rio de Janeiro, Conselho Nacional de Desenvolvimento Científico e Tecnológico and the Ministério da Ciência, Tecnologia e Inovação, the Deutsche Forschungsgemeinschaft and the Collaborating Institutions in the Dark Energy Survey.

  The Collaborating Institutions are Argonne National Laboratory, the University of California at Santa Cruz, the University of Cambridge, Centro de Investigaciones Enérgeticas, Medioambientales y Tecnológicas–Madrid, the University of Chicago, University College London, the DES-Brazil Consortium, the University of Edinburgh, the Eidgenössische Technische Hochschule (ETH) Zürich, Fermi National Accelerator Laboratory, the University of Illinois at Urbana-Champaign, the Institut de Ciències de l’Espai (IEEC/CSIC), the Institut de Física d’Altes Energies, Lawrence Berkeley National Laboratory, the Ludwig-Maximilians Universität München and the associated Excellence Cluster Universe, the University of Michigan, NSF NOIRLab, the University of Nottingham, the Ohio State University, the OzDES Membership Consortium, the University of Pennsylvania, the University of Portsmouth, SLAC National Accelerator Laboratory, Stanford University, the University of Sussex, and Texas A\&M University.

  Based on observations at NSF Cerro Tololo Inter-American Observatory, NSF NOIRLab (NOIRLab Prop. ID 2024B-261264; PI: Chow-Choong Ngeow), which is managed by the Association of Universities for Research in Astronomy (AURA) under a cooperative agreement with the U.S. National Science Foundation.

  This publication is based in part on proprietary Rubin Observatory Legacy Survey of Space and Time (LSST) data, and was prepared in accordance with the Rubin Observatory data rights and access policies. All authors of this publication meet the requirements for co-authorship of proprietary LSST data

  This material is based upon work supported in part by the National Science Foundation through Cooperative Agreements AST-1258333 and AST-2241526 and Cooperative Support Agreements AST-1202910 and 2211468 managed by the Association of Universities for Research in Astronomy (AURA), and the Department of Energy under Contract No. DE-AC02-76SF00515 with the SLAC National Accelerator Laboratory managed by Stanford University. Additional Rubin Observatory funding comes from private donations, grants to universities, and in-kind support from LSST-DA Institutional Members. This research uses services or data provided by the Rubin Science Platform at NSF-DOE Vera C. Rubin Observatory, which is jointly funded by the U.S. National Science Foundation and the U.S. Department of Energy, Office of Science.
  
\end{acknowledgments}



%

\facilities{Blanco (DECam), Rubin:Simonyi , Rubin:USDAC}

\software{{\tt Aladin} \citep{bonnarel2000}, {\tt astropy} \citep{astropy2013,astropy2018,astropy2022}, {\tt ds9} \citep{ds92026}, {\tt dustmaps} \citep{green2018}, {\tt MissFITS} \citep{marmo2008}, {\tt PSFEx} \citep{bertin2011}, {\tt Swarp} \citep{bertin2002}, {\tt Source-Extractor} \citep{bertin1996}}

\bibliography{DECamLMCMira_v1}{}
\bibliographystyle{aasjournal}



\end{document}